\documentclass[graybox]{svmult}

\usepackage{helvet}         
\usepackage{courier}        
\usepackage{type1cm}        
\usepackage{makeidx}         
\usepackage{graphicx}        
\usepackage{multicol}        
\usepackage[bottom]{footmisc}

\usepackage{cite}

\def\ii{{\rm i}}

\newcommand{\bbox}{\lower.2ex\hbox{$\Box$}}

\def\bfone{\relax{\rm 1\kern-.35em 1}}
\def\bfzero{\relax{\rm I\kern-.18em 0}}

\usepackage{amsmath}
\usepackage{amssymb}
\usepackage{amsfonts}
\usepackage{mathtools}
\usepackage{microtype}

\begin{document}

\title*{On the hidden symmetries of $D=11$ supergravity}
\author{Lucrezia Ravera}
\institute{Lucrezia Ravera \at DISAT, Politecnico di Torino, Corso Duca degli Abruzzi 24, 10129 Torino, Italy and INFN, Sezione di Torino, Via P. Giuria 1, 10125 Torino, Italy. \\ \email{lucrezia.ravera@polito.it}}

\maketitle

\abstract{We report on recent developments regarding the supersymmetric Free Differential
Algebra describing the vacuum structure of $D=11$ supergravity. We focus on the emergence of a
hidden superalgebra underlying the theory, explaining the group-theoretical role played
by the nilpotent fermionic generator naturally appearing for consistency of the construction. We also discuss the relation between this hidden superalgebra and other superalgebras of particular relevance in the context of supergravity and superstring, involving a fermionic generator with 32 components.}

\vspace{1.5cm}

\noindent
Contribution to the Proceedings of the XIV International Workshop ``Lie Theory and its Applications in Physics'', 20-25 June 2021, organized from Sofia, Bulgaria (on-line), based on a talk given by Lucrezia Ravera.


\section{Introduction}\label{s1}

In eleven spacetime dimensions an (almost) central extension of the supersymmetry algebra was introduced in the literature and named M-algebra \cite{deAzcarraga:1989mza,Hassaine:2003vq,Hassaine:2004pp,Sezgin:1996cj,Townsend:1997wg}. Such super Lie algebra includes, besides the super-Poincaré structure, the anticommutator
\begin{equation}
\lbrace Q , Q \rbrace = - \ii  \left(C
\Gamma ^{a} \right) P_{a} - \frac{1}{2} \left( C \Gamma
^{ab}\right) Z_{ab} - \frac{\ii}{5!}\left( C \Gamma ^{a_1\cdots a_5} \right)Z_{a_1\cdots a_5} \,, \label{malg}
\end{equation}
where $Z_{ab}$ and $Z_{a_1\cdots a_5}$ are Lorentz-valued almost central charges (they commute with all generators except the Lorentz one).
The M-algebra is commonly considered as the Lie superalgebra underlying M-theory \cite{Duff:1996aw,Schwarz:1995jq,Townsend:1996xj} in its low-energy limit, which corresponds to $D=11$ supergravity in the presence of non-trivial M2- or M5-brane sources \cite{Achucarro:1987nc,Bergshoeff:1987cm,Bergshoeff:1987qx,Duff:1987bx,Townsend:1995kk,Townsend:1995gp}. 
However, a field theory based on the M-algebra \eqref{malg} is naturally defined on a superspace that is enlarged with respect to the ordinary one. Indeed, let us recall that ordinary superspace is spanned by the supervielbein $\lbrace V^a , \Psi \rbrace$, where $V^a$ is the bosonic vielbein and $\Psi$ the gravitino 1-form, while the base space induced by the M-algebra includes also the bosonic 1-form fields $B^{ab}$ and $B^{a_1\cdots a_5}$, respectively dual to the generators $Z_{ab}$ and $Z_{a_1 \cdots a_5}$. On the other hand, the low-energy limit of the M-theory, corresponding to $D=11$ supergravity, should be based on ordinary superspace. Under this perspective, the M-algebra cannot be the final answer, as it is not sufficient to reproduce the Free Differential Algebra \cite{Sullivan} (FDA in the following) on which $D=11$ supergravity is based. Here let us recall that $D=11$ supergravity \cite{Cremmer:1978km} contains, besides the super-Poincaré fields given by the Lorentz spin connection $\omega^{ab}$ and the supervielbein $\lbrace V^a , \Psi \rbrace$, with $a,b,\ldots = 0, 1, \ldots , 10$, also a 3-form $A^{(3)}$, satisfying, in the superspace vacuum,
\begin{equation}
dA^{(3)} - \frac{1}{2} \bar{\Psi} \wedge \Gamma_{ab} \Psi \wedge V^a \wedge V^b = 0 \,,
\end{equation}
whose closure relies on 3-fermion 1-forms Fierz identities in superspace. As such, this theory is not based on a superalgebra, but instead on a FDA on the superspace spanned by the supervielbein.

A super Lie algebra of 1-forms leaving invariant $D=11$ supergravity and reproducing the FDA on ordinary superspace was introduced in 1981 by R. D'Auria and P. Fré in \cite{D'Auria:1982nx} and later named D'Auria-Fré algebra (DF-algebra in the following).\footnote{The DF-algebra
has recently raised a certain interest also in the Mathematical-Physicists community, due to the fact
that it can be reformulated in terms of $\mathcal{L}_n \subset \mathcal{L}_{\infty}$ algebras, or ``strong homotopy Lie algebras'', see, e.g., \cite{Sati:2015yda}.} Such Lie superalgebra, from which the M-algebra emerges as a subalgebra, includes, besides the Lorentz ($J_{ab}$), spacetime translations ($P_a$) and supersymmetry ($Q_\alpha$, $\alpha=1,\ldots,32$) generators, and the bosonic charges $Z_{ab}$, $Z_{a_1 \cdots a_5}$, also a nilpotent fermionic charge $Q'$ ($\lbrace Q',Q'\rbrace =0$), such that
\begin{equation}
[P_a,Q] \propto \Gamma_a Q' \,, \quad [Z_{ab},Q] \propto \Gamma_{ab} Q' \,, \quad [Z_{a_1 \cdots a_5},Q] \propto \Gamma_{a_1 \cdots a_5} Q' \,. \label{commu}
\end{equation}
The presence of the extra fermionic nilpotent charge $Q'$ in the D'Auria-Fré construction is naturally required by supersymmetry and consistency of the theory. Let us stress that, actually, this fact is not a peculiarity of $D=11$ supergravity, but is fully general: a hidden superalgebra underlying the supersymmetric FDA containing at least one nilpotent fermionic generator can be constructed for any supergravity theory involving antisymmetric tensor fields \cite{Andrianopoli:2016osu}. 

It was hence proven that the hidden superalgebra underlying the FDA of $D=11$ supergravity is the DF-algebra, in the sense that it is equivalent to the FDA description of the $D=11$ theory on ordinary superspace (and therefore to the Cremmer-Julia-Scherk theory \cite{Cremmer:1978km}). The bosonic generators $Z_{ab}$ and $Z_{a_1 \cdots a_5}$ were later understood as $p$-brane charges, sources of the dual potentials $A^{(3)}$ and $B^{(6)}$ appearing in the theory, and eq. \eqref{malg} was interpreted as the natual generalization of the supersymmetry algebra in higher dimensions, in the presence of non-trivial topological extended sources (black $p$-branes). On the other hand, a clearer understanding of the group-theoretical and physical meaning of the (necessary) nilpotent fermionic generator $Q'$ has been provided only reather recently, in \cite{Andrianopoli:2016osu}.
Besides, some other issues remained open, such as: Is the DF-algebra the fully extended superalgebra underlying $D=11$ supergravity? And would a non-Abelian charge deformation of the DF-algebra be possible? Which is the relation between the DF-algebra and the most general simple superalgebra involving a fermionic generator with 32 components, namely $\mathfrak{osp}(1|32)$? An answer to these questions was formulated in the works \cite{Andrianopoli:2016osu,Andrianopoli:2017itj}, as we are going to review in the following. Before proceeding in this direction, let us briefly recap some key aspects of the geometric approach to supergravity in superspace \cite{CDF} adopted and the FDA description of $D=11$ supergravity.

\section{Lie superalgebras and Maurer-Cartan equations}\label{s2}

In the geometric approach to supergravity in superspace \cite{CDF} the dual formulation of Lie superalgebras in terms of the associated Maurer-Cartan (MC) equations is adopted: Given a Lie superalgebra
\begin{equation}
[T_A , T_B \rbrace = {C_{AB}}^C T_C \,,
\end{equation}
where $T_A$ are the generators in the adjoint representation of the corresponding Lie supergroup, one can introduce an equivalent description in terms of the differential 1-forms $\sigma^A$ dual to the Lie superalgebra generators, $\sigma^A (T_B) = {\delta^A}_B$, obeying the MC equations
\begin{equation}
R^A \equiv d \sigma^A + \frac{1}{2} {C_{BC}}^A \sigma^B \wedge \sigma^C = 0 \,. \label{mc}
\end{equation}
In order to describe non-trivial physical configurations, a non-vanishing right-hand side has to be switched on in \eqref{mc}, which corresponds to defining the supercurvatures $R^A$ (super field-strengths). The latter are the building blocks of supergravity in the geometric approach.
The MC equations $R^A=0$ can be identified with the vacuum configuration of a supergravity theory, and their $d^2$-closure is equivalent to the Jacobi identities of the dual algebraic structure. Moreover, as we are dealing with the geometric formulation in superspace, let us stress that the latter is spanned by the supervielbein $\lbrace V^a , \Psi \rbrace$, the 1-form fields $V^a$ and $\Psi$ being respectively dual to the generators $P_a$ and $Q$. For a detailed review of the the geometric approach to supergravity in superspace we refer the reader to \cite{DAuria:2020guc}.

\section{$D=11$ supergravity and its Free Differential Algebra}\label{s3}

Supergravity theories in $4\leq D \leq 11$ spacetime dimensions have a bosonic field content that generically includes, besides the metric and a set of 1-form gauge potentials,
also $p$-index antisymmetric tensors, and they are therefore appropriately discussed in the FDAs framework. 
Indeed, FDAs extend the MC equations by incorporating $p$-form gauge potentials. Let us schematically review the steps for constructing a FDA: Given a set of MC 1-forms $\lbrace \sigma^A \rbrace$, we can build up $n$-form cochains,
\begin{equation}
\Omega^{(n)} = \Omega_{A_1 \cdots A_n} \sigma^{A_1} \wedge \cdots \wedge \sigma^{A_n} \,.
\end{equation}
Then, if there exists a $p$ such that $d\Omega^{(p+1)}=0$, i.e., a cocycle, we can introduce a $p$-form (gauge potential) $A^{(p)}$ such that
\begin{equation}
F^{(p+1)} \equiv dA^{(p)} + \Omega^{(p+1)} = 0 \,.
\end{equation}
Consequently, we can consider $\left(\lbrace \sigma^A \rbrace , A^{(p)} \right)$ as new a basis of MC forms and look for new cocycles, iteratively, constructing the complete FDA.

We now turn to the FDA description of the (vacuum structure of) $D=11$ supergravity. The theory, which in particular involves a 3-index antisymmetric tensor $A_{\mu \nu \rho}$ ($\mu, \nu, \rho, \ldots = 0,1,\ldots,10$), was originally built in 1978 \cite{Cremmer:1978km} and subsequently reformulated geometrically by R. D'Auria and P. Fré in \cite{D'Auria:1982nx} in terms of a supersymmetric FDA on superspace. The latter reads as follows:
\begin{eqnarray}
\mathcal{R}^{ab} &=& d\omega^{ab} - \omega^{ac} \wedge {\omega_c}^b = 0 \,, \nonumber \\
R^a &\equiv& \mathcal{D} V^a - \frac{\ii}{2} \bar{\Psi} \wedge \Gamma^a \Psi = 0 \,, \nonumber \\
\rho &\equiv& \mathcal{D} \Psi = 0 \,, \nonumber \\
F^{(4)} &\equiv& dA^{(3)} - \frac{1}{2} \bar{\Psi} \wedge \Gamma_{ab} \Psi \wedge V^a \wedge V^b = 0 \,, \label{FDA11}
\end{eqnarray}
with $A^{(3)}=A_{\mu \nu \rho}dx^\mu \wedge dx^\nu \wedge dx^\rho$ and where $\mathcal{D}=d-\omega$ denotes the Lorentz covariant derivative. The right-hand side of \eqref{FDA11} defines the vacuum of the theory. The $d^2$-closure of the last equation in \eqref{FDA11} relies on the $3\Psi$ Fierz identity
\begin{equation}
\Gamma_{ab} \Psi \wedge \bar{\Psi} \wedge \Gamma^a \Psi = 0 \,.
\end{equation}
Furthermore, due to another $3\Psi$ Fierz identity, that is
\begin{equation}
\Gamma_{[a_1 a_2} \Psi \wedge \bar{\Psi} \wedge \Gamma_{a_3 a_4]} \Psi + \frac{1}{3} \Gamma_{a_1 \cdots a_5} \Psi \wedge \bar{\psi} \wedge \Gamma^{a_1 \cdots a_5} \Psi = 0 \,,
\end{equation}
the supersymmetric FDA also allows to include in the description
\begin{equation}
F^{(7)} \equiv dB^{(6)} - 15 A^{(3)} \wedge dA^{(3)} - \frac{\ii}{2} \bar{\Psi} \wedge \Gamma_{a_1 \cdots a_5} \Psi \wedge V^{a_1} \wedge \cdots \wedge V^{a_5} = 0 \,, \label{FDA11b}
\end{equation}
$F^{(7)}$ being Hodge-dual to $F^{(4)}$ on spacetime. The complete FDA is therefore defined in terms of $\left( V^a, \Psi , A^{(3)}, B^{(6)} \right)$, and it is invariant under the $p$-form gauge transformations
\begin{equation}
\delta A^{(3)} = d \Lambda^{(2)} \,, \quad \delta B^{(6)} = d \Lambda^{(5)} + 15 \Lambda^{(2)} \wedge d A^{(3)} \,, \label{gaugetr}
\end{equation}
with $p$-form gauge parameters $\Lambda^{(2)}$ and $\Lambda^{(5)}$. 

\section{Hidden superalgebra underlying $D=11$ supergravity}\label{s4}

The investigation presented in \cite{D'Auria:1982nx} proved that the FDA reported in the previous section can be traded for an ordinary Lie superalgebra. The D'Auria-Fré recipe consists of the following steps:
\begin{enumerate}
\item Associate to $A^{(3)}$ and $B^{(6)}$ the 1-form fields $B^{ab}=B^{[ab]}$ and $B^{a_1 \cdots a_5}=B^{[a_1 \cdots a_5]}$, respectively;
\item Take as basis of MC 1-forms $\sigma^A \equiv \lbrace V^a, \Psi, \omega^{ab}, B^{ab}, B^{a_1 \cdots a_5} \rbrace$, which implies the extra MC equations
\begin{equation}
\mathcal{D} B^{ab} = \frac{1}{2} \bar{\Psi} \wedge \Gamma^{ab} \Psi \,, \quad \mathcal{D} B^{a_1 \cdots a_5} = \frac{\ii}{2} \bar{\psi} \wedge \Gamma^{a_1 \cdots a_5} \Psi \,;
\end{equation}
\item Assume $A^{(3)}$ to be written in terms of the 1-forms $\sigma^A$ above, that is $A^{(3)}=A^{(3)}(\sigma)$, with all possible combinations,
\begin{eqnarray}
A^{(3)}(\sigma) &=& T_0 B_{ab} \wedge V^a \wedge V^b + T_1 B_{a b}\wedge {B^{b}}_{c}\wedge B^{c a} \\
&& + T_2 B_{b_1 a_1 \cdots a_4}\wedge {B^{b_1}}_{b_2}\wedge B^{b_2 a_1 \cdots a_4} \nonumber \\
&& + T_3 \epsilon_{a_1 \cdots a_5 b_1 \cdots b_5 m}B^{a_1 \cdots a_5}\wedge B^{b_1 \cdots b_5}\wedge V^m \nonumber \\
&& + T_4 \epsilon_{m_1 \cdots m_6 n_1 \cdots n_5}B^{m_1 m_2 m_3 p_1 p_2}\wedge {B^{m_4 m_5 m_6}}_{p_1 p_2}\wedge B^{n_1 \cdots n_5} \,; \nonumber
\end{eqnarray}
where $T_0$, $T_1$, $T_2$, $T_3$, $T_4$ are constant parameters;
\item Require
\begin{equation}
dA^{(3)} (\sigma) = \frac{1}{2} \bar{\Psi} \wedge \Gamma_{ab} \Psi \wedge V^a \wedge V^b \,, \label{req}
\end{equation}
namely that the vacuum FDA structure on ordinary superspace, spanned by the supervielbein $\lbrace V^a, \Psi \rbrace$, is reproduced once considering $A^{(3)}=A^{(3)} (\sigma)$.
\end{enumerate}
Then, expressing the FDA with $A^{(3)}=A^{(3)} (\sigma)$ determines the latter expression. However, as shown in \cite{D'Auria:1982nx}, this requires to include in the parametrization of $A^{(3)}$ in terms of 1-forms a spinor 1-form field $\eta$, such that
\begin{equation}
\mathcal{D} \eta = \ii E_1 \Gamma_a \Psi \wedge V^a + E_2 \Gamma_{ab} \Psi \wedge B^{ab} + \ii E_3 \Gamma_{a_1 \cdots a_5} \Psi \wedge B^{a_1 \cdots a_5} \,,
\end{equation}
whose $d^2$-closure requires $E_1 + 10 E_2 - 5! E_3 = 0$, enlarging in this way the basis of MC 1-forms to $\sigma^A \equiv \lbrace V^a, \Psi, \omega^{ab}, B^{ab}, B^{a_1 \cdots a_5}, \eta \rbrace$. We are therefore led to consider
\begin{eqnarray}
A^{(3)}(\sigma) &=& T_0 B_{ab} \wedge V^a \wedge V^b + T_1 B_{a b}\wedge {B^{b}}_{c}\wedge B^{c a} \\
&& + T_2 B_{b_1 a_1 \cdots a_4}\wedge {B^{b_1}}_{b_2}\wedge B^{b_2 a_1 \cdots a_4} \nonumber \\
&& + T_3 \epsilon_{a_1 \cdots a_5 b_1 \cdots b_5 m}B^{a_1 \cdots a_5}\wedge B^{b_1 \cdots b_5}\wedge V^m \nonumber \\
&& + T_4 \epsilon_{m_1 \cdots m_6 n_1 \cdots n_5}B^{m_1 m_2 m_3 p_1 p_2}\wedge {B^{m_4 m_5 m_6}}_{p_1 p_2}\wedge B^{n_1 \cdots n_5} \nonumber \\
&& + \ii S_1 \bar{\Psi}\wedge \Gamma_a \eta \wedge V^a + S_2 \bar{\Psi}\wedge \Gamma_{ab} \eta \wedge B^{ab} \nonumber \\
&& + \ii S_3 \bar{\Psi}\wedge \Gamma_{a_1 \cdots a_5} \eta \wedge B^{a_1 \cdots a_5} \,, \nonumber
\end{eqnarray}
and the requirement \eqref{req} fixes the coefficients $E_i$, $T_j$, $S_k$ in terms of a single free parameter \cite{Bandos:2004xw}. The dual set of generators spanning the hidden superalgebra underlying $D=11$ supergravity (i.e., the DF-algebra) is $T_A \equiv \lbrace P_a , Q, J_{ab}, Z_{ab}, Z_{a_1 \cdots a_5}, Q' \rbrace$. The DF-algebra includes, besides the super-Poincaré structure, the anticommutation relations \eqref{malg} and $\lbrace Q',Q' \rbrace =0$, and the commutators \eqref{commu}. As we have anticipated before, the generators $Z_{ab}$ and $Z_{a_1 \cdots a_5}$ were later understood as M-brane charges, sources of $A^{(3)}$ and $B^{(6)}$, respectively, while the role played by the necessary nilpotent fermionic charge $Q'$ was clarified in \cite{Andrianopoli:2016osu} (and \cite{Andrianopoli:2017itj}). We report on this in the following.

\subsection{Role of the nilpotent fermionic generator $Q'$}\label{ss1}

The DF-algebra is a ``spinorial central extension'' of the M-algebra including $Q'$. In \cite{Andrianopoli:2016osu} it was shown that the inclusion of the spinor 1-form $\eta$ (dual to $Q'$), whose presence in naturally required by supersymmetry in the D'Auria-Fré construction, allows to realize the M-algebra as a (hidden) symmetry of $D=11$ supergravity. In particular, $\eta$ allows a fiber bundle structure $\mathcal{G} \rightarrow \text{(superspace)}$ on the group-manifold $\mathcal{G}$ generated by the M-algebra, intertwining between basis and fiber. The $p$-form gauge transformations leaving invariant the supersymmetric FDA trivialized in terms of 1-forms result to be realized as diffeomorphisms in the fiber direction of the group-manifold $\mathcal{G}$.

More precisely, if $\eta \neq 0$, the hidden superalgebra, let us call it $\mathbb{G}$, generates a group-manifold $\mathcal{G}$ with a principal fiber bundle structure $\mathcal{G} \rightarrow K$, where the base space $K$ is superspace, spanned by $\lbrace V^a, \Psi \rbrace \in \mathbb{K}$, and we have
\begin{equation}
dA^{(3)} (\sigma) = \frac{1}{2} \bar{\Psi} \wedge \Gamma_{ab} \Psi \wedge V^a \wedge V^b  \in \mathbb{K} \times \cdots \times \mathbb{K} \,,
\end{equation}
while the fiber is generated by $\mathbb{H}=H_0 + \mathcal{H}$, where
\begin{equation}
\lbrace \omega^{ab} \rbrace \in H_0 \,, \quad \lbrace B^{ab} , B^{a_1 \cdots a_5} \rbrace \in \mathcal{H} \,.
\end{equation}
The spinor 1-form $\eta$ behaves like a cohomological ``ghost'' field, in the sense that it allows to realize in a dynamical way the gauge invariance of $A^{(3)}$, guaranteeing that only the physical degrees of freedom appear in the FDA (namely that the FDA on ordinary superspace is reproduced). A singular limit $\eta \rightarrow 0$ exits, where a trivialization $A^{(3)}_{\text{lim}}(\sigma)$ can still be defined with the same $\mathcal{G}$ but $dA^{(3)}_{\text{lim}}(\sigma) \in \mathbb{G} \times \cdots \times \mathbb{G}$, namely the FDA with $\eta \rightarrow 0$ lives in an enlarged superspace.

Concerning gauge invariance, for $A^{(3)}(\sigma)$ the $p$-form gauge transformations of the FDA are realized through gauge transformations in $\mathcal{H}$,
\begin{equation}
\begin{cases}
\delta B^{ab} = \mathcal{D} \Lambda^{ab} \,, \\
\delta B^{a_1 \cdots a_5} = \mathcal{D} \Lambda^{a_1 \cdots a_5} 
\end{cases}
\quad \Rightarrow \quad
\begin{cases}
\delta A^{(3)} = d \Lambda^{(2)} \,, \\
\delta B^{(6)} = d \Lambda^{(5)} + 15 \Lambda^{(2)} \wedge d A^{(3)} \,.
\end{cases}
\end{equation}
The gauge invariance of the FDA trivialized in terms of 1-forms requires
\begin{equation}
\delta_{\text{gauge}} \eta = - E_2 \Lambda^{ab} \Gamma_{ab} \Psi - \ii E_3 \Lambda^{a_1 \cdots a_5} \Gamma_{a_1 \ldots a_5} \Psi \,.
\end{equation}
Hence, considering the tangent vector 
\begin{equation}
\overrightarrow{z} \equiv \Lambda^{ab}Z_{ab} + \Lambda^{a_1 \cdots a_5}Z_{a_1 \cdots a_5}
\end{equation}
in $\mathcal{H} \in \mathbb{G}$, we find that there exists a $\bar{\Lambda}^{(2)} = \Lambda^{(2)}(\Lambda^{ab}, \Lambda^{a_1 \cdots a_5};\sigma) = \imath_{\overrightarrow{z}} \left(A^{(3)}(\sigma)\right)$, where $\imath$ denotes the contraction operator, such that
\begin{equation}
\delta_{\bar{\Lambda}} \left( A^{(3)} (\sigma) \right) = d \bar{\Lambda}^{(2)} = \ell_{\overrightarrow{z}} \left(A^{(3)}(\sigma)\right) \,, \label{gaugetrlieder}
\end{equation}
where $\ell_{\overrightarrow{z}} = d \imath_{\overrightarrow{z}} + \imath_{\overrightarrow{z}} d$ is the Lie derivative in the direction $\overrightarrow{z}$ and where we have also used the fact that $\imath_{\overrightarrow{z}} \left(dA^{(3)}\right)=0$.
Therefore, for $\eta \neq 0$, $\delta A^{(3)}$ is genuinely realized as a diffeomorphism in the fiber direction of the group-manifold $\mathcal{G}$. Let us finally mention that, as shown in \cite{Andrianopoli:2016osu}, assuming $\bar{\Lambda}^{(5)}= \imath_{\overrightarrow{z}} \left(B^{(6)} (\sigma)\right)$ one can prove that $\delta_{\bar{\Lambda}} B^{(6)} = \ell_{\overrightarrow{z}} \left(B^{(6)}(\sigma)\right)$, whatever $B^{(6)}(\sigma)$ would be. This is particularly relevant since, even though we do not know the explicit parametrization of $B^{(6)}$ in terms of 1-form fields, at least we can say that, analogously to what happens for $A^{(3)}$, $\delta B^{(6)}$ is properly realized as a diffeomorphism in the fiber direction of $\mathcal{G}$. Remarkably, as the structure above relies on supersymmetry (and, in particular, on $3\Psi$ Fierz identities), the extension of this analysis to lower dimensions might turn out to be a useful tool in generalized geometry frameworks, such as Exceptional Field Theory, offering a dynamical way to implement the so-called section constraints.

One might ask at this point whether the DF-algebra is the fully extended superalgebra underlying $D=11$ supergravity.
Some clues to answer this question are provided to us by minimal $D=7$ supergravity, in which case the full on-shell hidden symmetry involves two nilpotent fermionic charges, associated with the presence of two mutually dual $p$-forms \cite{Andrianopoli:2016osu}. It could therefore be conjectured that there are different spinors associated with the mutually dual $p$-forms even in $D=11$, and that new extra 1-forms may be necessary to write the parameterization $B^{(6)}(\sigma)$ in such a way to reproduce the complete FDA on ordinary superspace. Under this perspective, it would be particularly useful to calculate explicitly $B^{(6)}(\sigma)$.

\section{Relation between the DF-algebra and $\mathfrak{osp}(1|32)$}\label{s5}

To come to more realistic cases, it would be important to be able to switch on non-Abelian charges in the setup reviewed in the previous sections. Such issue could be analyzed either as in Exceptional Field Theory, by Scherk-Schwarz dimensional reduction to lower dimensions, or directly in $D=11$. However, it is well-known that in $D=11$ the massive theory is problematic. Nevertheless, let us have a look closer, reporting the results obtained in \cite{Andrianopoli:2017itj} pointing in the direction of a clearer understanding of the problem. In fact, as we have previously mentioned, the DF-algebra and the parametrization $A^{(3)}(\sigma)$ depend on a free parameter. In \cite{Andrianopoli:2017itj} it was shown that this dependence can be associated with an intriguing relation with $\mathfrak{osp}(1|32)$, which is the most general simple superalgebra involving a fermionic generator with 32 components and a scale parameter $e$. The latter has length dimension $-1$ and can be thought as proportional to (the square root of) a cosmological constant.

The 1-form fields appearing in the dual formulation of $\mathfrak{osp}(1|32)$ are $\lbrace V^a, \Psi, \omega^{ab}, B^{a_1 \cdots a_5} \rbrace$. To make contact with the DF-algebra, it is first of all necessary to include a further bosonic 1-form field $B^{ab}$, and this was done in \cite{Castellani:1982kd} by considering a ``torsion deformation'' of $\mathfrak{osp}(1|32)$, namely taking
\begin{equation}
\omega^{ab} \rightarrow \omega^{ab} - e B^{ab} \,, \quad \mathcal{R}^{ab} \rightarrow \mathcal{R}^{ab} - e \mathcal{D} B^{ab} + e^2 B^{ac} \wedge {B_c}^b 
\end{equation}
and then requiring a Minkowski background, $\mathcal{R}^{ab} \equiv d\omega^{ab} - \omega^{ac} \wedge {\omega_c}^b = 0$. Moreover, in order to try to make contact also with $\eta$ of the DF-algebra, in \cite{Castellani:1982kd} such torsion deformation of $\mathfrak{osp}(1|32)$ was enlarged by including a spinor 1-form $\eta^e$. The MC description of the resulting superalgebra reads
\begin{eqnarray}
\mathcal{R}^{ab} &\equiv& d\omega^{ab} -\omega^{ac}\wedge {\omega_{c}}^{b} = 0 \,, \label{tordefosp} \\
\mathcal{D}V^{a} &=& - e B^{ab} \wedge V_b + \frac{e}{2\cdot (5!)^2}\epsilon^{a b_1 \cdots b_5 c_1 \cdots c_5}B_{b_1 \cdots b_5}\wedge B_{c_1 \cdots c_5} \nonumber \\
&& + \frac{\ii}{2}\bar{\Psi} \wedge \Gamma^a \Psi \,, \nonumber  \\
\mathcal{D} B^{ab} &=& e V^a \wedge V^b - e B^{ac}\wedge {B_{c}}^{b}+ \frac{e}{24}B^{a b_1 \cdots b_4}\wedge {B^b}_{b_1 \cdots b_4} + \frac{1}{2}\bar{\Psi}\wedge \Gamma^{ab}\Psi \,, \nonumber \\
\mathcal{D} B^{a_1 \cdots a_5} &=& 5 e B^{m [a_1}\wedge {B^{a_2 \cdots a_5 ]}}_{m} + \frac{e}{5!}\epsilon^{a_1 \cdots a_5 b_1 \ldots c_6}B_{b_1 \cdots b_5}\wedge V_{b_6} \nonumber \\
&& -\frac{5 e}{6!}\epsilon^{a_1 \cdots a_5 b_1 \cdots b_6}{B^{c_1 c_2}}_{b_1 b_2 b_3}\wedge B_{c_1 c_2 b_4 b_5 b_6}+ \frac{\ii}{2}\bar{\Psi}\wedge \Gamma^{a_1 \cdots a_5}\Psi \,, \nonumber \\
\mathcal{D} \Psi &=& \frac{\ii}{2} e \Gamma_a \Psi \wedge V^a + \frac{1}{4}e \Gamma_{ab}\Psi \wedge B^{ab}+ \frac{\ii}{2 \cdot 5!} e \Gamma_{a_1 \cdots a_5}\Psi \wedge B^{a_1 \cdots a_5} \,, \nonumber \\
\mathcal{D} \eta^e &=& \frac{\ii}{2} \Gamma_a \psi \wedge V^a + \frac{1}{4} \Gamma_{ab}\Psi \wedge B^{ab}+ \frac{\ii}{2 \cdot 5!} \Gamma_{a_1 \cdots a_5}\Psi \wedge B^{a_1 \cdots a_5} \,. \nonumber
\end{eqnarray}
In particular, we can see that $\mathcal{D} \eta^e=\frac{1}{e} \mathcal{D}\Psi$, and the MC closure does not allow any free parameters. Hence, in the $e \rightarrow 0$ limit the algebraic structure above does not reproduce the DF-algebra, as $\eta^e_\neq \eta$ for any value of the free parameter in the DF-algebra. On the other hand, let us observe that the reduced version of \eqref{tordefosp} without $\eta^e$ in the limit $e \rightarrow 0$ gives the M-algebra.

The relation between the DF-algebra and $\mathfrak{osp}(1|32)$ was subsequently clarified in \cite{Andrianopoli:2017itj} under a cohomological perspective. In particular, in \cite{Andrianopoli:2017itj} it was found that $A^{(3)}(\sigma)$ and $\eta$ admit the general decomposition
\begin{equation}
A^{(3)}(\sigma) = A^{(3)}_{(0)} + \alpha A^{(3)}_{(e)} \,, \quad \eta = \eta_{(0)} + \alpha \eta_{(e)} \,, \label{decomp}
\end{equation}
$\alpha$ being precisely the free parameter of the D'Auria-Fré construction. The contribution
\begin{equation}
A^{(3)}_{(0)}=A^{(3)}_{(0)}(V^a,\Psi,B^{ab},\eta_{(0)})
\end{equation}
does not depend on $B^{a_1 \cdots a_5}$ and explicitly breaks the $\mathfrak{osp}(1|32)$ structure, while the contribution
\begin{equation}
A^{(3)}_{(e)} = A^{(3)}_{(e)} (V^a,\Psi,B^{ab},B^{a_1 \cdots a_5},\eta_{(e)})
\end{equation}
is covariant under the (torsion deformation of) $\mathfrak{osp}(1|32)$. In the vacuum FDA we have
\begin{equation}
dA^{(3)}_{(0)} = \frac{1}{2} \bar{\Psi} \wedge \Gamma_{ab} \Psi \wedge V^a \wedge V^b \,, \quad dA^{(3)}_{(e)} = 0 \,.
\end{equation}
Therefore, it emerges that only $dA^{(3)}_{(0)}$ is responsible for the 4-form cohomology of the supersymmetric FDA, and the free parameter $\alpha$ parametrizes the cohomologically trivial deformation $dA^{(3)}_{(e)}$.
We conclude that, as the decomposition \eqref{decomp} shows, $A^{(3)}(\sigma)$ is not invariant under $\mathfrak{osp}(1|32)$ (neither under its torsion deformation) because of the contribution $A^{(3)}_{(0)}$ explicitly breaking this
symmetry. Such term is however the only one contributing to the vacuum 4-form cohomology in superspace.

It would be worth conducting a similar analysis for the 6-form $B^{(6)}$. Furthermore, one might then consider the out-of-vacuum FDA, where we would have
\begin{equation}
dA^{(3)} - \frac{1}{2} \bar{\Psi} \wedge \Gamma_{ab} \Psi \wedge V^a \wedge V^b = F^{(4)} = F^{(4)}_{(0)} + \alpha F^{(4)}_{(e)}
\end{equation}
with
\begin{equation}
dA^{(3)}_{(0)} = \frac{1}{2} \bar{\Psi} \wedge \Gamma_{ab} \Psi \wedge V^a \wedge V^b + F^{(4)}_{(0)} \,, \quad dA^{(3)}_{(e)} = F^{(4)}_{(e)} \,,
\end{equation}
and compute the charge associated with the 3-form gauge potential,
\begin{equation}
q = \int dA^{(3)} = q_{(0)} + \alpha q_{(e)} \,.
\end{equation}
In this context, possible connections could emerge with the analysis of the 4-form cohomology of M-theory on spin manifolds \cite{Witten:1996md,Diaconescu:2000wy}, where $dA^{(3)}_{(0)}$ might turn out to be the contribution responsible for the canonical integral class of the spin bundle of $D=11$ superspace. In particular, this would imply that $q_{(0)}$ could assume fractional values (in units of $q_{(e)}$).

To conclude, let us also mention that the 4-form $F^{(4)}$ appears in the topological term $A^{(3)} \wedge F^{(4)} \wedge F^{(4)}$ of the $D=11$ supergravity Lagrangian, and it appears that the nilpotent spinor 1-form $\eta$ could be an important addition towards the construction of a possible off-shell theory underlying $D=11$ supergravity. In \cite{Hassaine:2003vq}, a supersymmetric $D=11$ Lagrangian invariant under the M-algebra and closing off-shell without requiring auxiliary fields was constructed, as a Chern-Simons form, and shown to depend on one free parameter. It would be very interesting to investigate the possible connections between this and the approach reviewed in the present report.

\begin{acknowledgement}
The author would like to thank the Department of Applied Science and Technology of the Polytechnic of Turin, and in particular Laura Andrianopoli and Francesco Raffa, for financial support. The work reported in this proceeding was performed in collaboration with Laura Andrianopoli and Riccardo D'Auria.
\end{acknowledgement}

\end{document}